\documentclass[useAMS,usenatbib]{mn2e}

\usepackage[psamsfonts]{amssymb}
\usepackage[dvips]{graphicx}
\usepackage{amsmath,alltt}
\usepackage{multirow}
\usepackage{rotating}
\usepackage{lscape}

\title[When does a star cluster become a multiple star system?]{When does a star cluster become a multiple star system? I. Lifetimes of equal-mass small-N systems}
\author[Leigh N. W. C., Shara M. M., Geller A. M.]{Nathan W. C. Leigh$^{1}$, Michael M. Shara$^{1}$, Aaron M. Geller$^{2,3}$
\thanks{E-mail: nleigh@amnh.org (NWCL), mshara@amnh.org, a-geller@northwestern.edu}\\
$^{1}$Department of Astrophysics, American Museum of Natural History, Central Park West and 79th Street, New York, NY 10024 \\
$^{2}$Center for Interdisciplinary Exploration and Research in Astrophysics (CIERA) and Department of Physics and Astronomy, \\ Northwestern University,\
 2145 Sheridan Rd, Evanston, IL 60208, USA \\
$^{3}$Adler Planetarium, Dept.\ of Astronomy, 1300 S. Lake Shore Drive, Chicago, IL 60605, USA}
\begin{document}

\pagerange{\pageref{firstpage}--\pageref{lastpage}} \pubyear{2011}

\maketitle

\label{firstpage}

\begin{abstract}
What is the difference between a long-lived unstable (or quasi-stable) multiple star system and a bona fide star cluster?  In this paper, we present a possible framework to address this question, by studying the distributions of disruption times for chaotic gravitational encounters as a function of the number of interacting particles.  
To this end, we perform a series of numerical scattering experiments with the \texttt{FEWBODY} code, to calculate the distributions of disruption times as a function of both the particle number N and the virial coefficient k.  The subsequent distributions are fit with a physically-motivated function, consisting of an initial exponential decay followed by a very slowly decreasing tail at long encounter times due to long-lived quasi-stable encounters.
We find three primary features characteristic of the calculated distributions of disruption times.  These are:  (1) the system half-life increases with increasing particle number, (2) the fraction of long-lived quasi-stable encounters increases with increasing particle number and (3) both the system half-life and the fraction of quasi-stable encounters increase with decreasing virial coefficient.  We discuss the significance of our results for collisional dynamics, and consider the extrapolation of our results to larger-N systems.  We suggest that this could potentially offer a clear and unambiguous distinction between star clusters and (unstable or quasi-stable) multiple star systems.  Although we are limited by very small-number statistics, our results tentatively suggest that (for our assumptions) this transition occurs at a critical particle number of order 100.
\end{abstract}


\section{Introduction} \label{intro}

The three-body problem has a long history extending all the way back to \citet{newton1686}.  
For the majority of the relevant parameter space and over sufficiently long timescales, the 
evolution is chaotic \citep{poincare1892}.  With the introduction of computers, it became possible to integrate the equations of motion directly using brute force, and a number of interesting behaviors characteristic of the chaotic three-body problem became apparent.  The system typically evolves via a series of close triple encounters  \citep[e.g.][]{agekyan67,anosova69,szebehely72,saslaw74,valtonen75,agekyan83,anosova83,anosova86,anosova94}.  Between each such event, one of the objects is temporarily ejected but remains bound to the three-body system.  This object recoils some distance from the remaining binary before returning to initiate another triple encounter.  Eventually, one of the bodies is ejected with a sufficiently high velocity to become unbound, and it escapes to infinity.  The total duration of this chaotic interplay is called the \textit{lifetime} of the bound system.

Interestingly, the distribution of disruption times $\tau_{\rm d}$ that follows from many escape trials, with a given escape probability per trial, is exponential \citep{valtonen75}.  Consequently, it can be described according to a half-life formalism, in analogy with radioactive decay.  Here, the half-life $\tau_{\rm 1/2}$ is defined as the time when the probability that the system remains intact is 50\%.  The probability of escape, and hence the system half-life, depends on a number of parameters, including the total encounter energy, total angular momentum as well as the mass of the escaper \citep[e.g.][]{valtonen06}.  The system half-life tends to increase with increasing angular momentum and also with an increasing virial coefficient, defined as the ratio of the total kinetic energy T to the total gravitational potential energy U, or k $=$ T/$|$U$|$ \citep[e.g.][]{anosova86,anosova94}.  Systems with a large range of masses tend to break up faster, since at least one of the interacting bodies has a low fraction of the total system mass and hence a high probability of escape \citep{szebehely72,anosova94}.  

Despite this early progress, many aspects of the lifetime statistics of interacting three-body systems remain poorly understood.  For example, on asymptotically long time-scales, exponential decay is in direct conflict with the theoretical prediction that the mean lifetime of an isolated three-body system is infinite \citep{agekyan83}.  More recently, \citet{mikkola07} revisited the distribution of disruption times for equal-mass three-body systems with randomized initial conditions.  As shown by \citet{shevchenko10} and \citet{orlov10}, the tails of the lifetime distributions are algebraic instead of exponential, and the differential distributions are better fit with a power-law index $\sim$ -3/2 (the exact value depends on the virial coefficient k).  

This suggests that the tail of the disruption time distribution should be considered separately from the initial exponential part, since it corresponds to a different dynamical state.  The physical mechanism responsible for this algebraic behavior is still under debate, but \citet{orlov10} speculate that the dynamical behavior at very long integration times is dominated by L\'{e}vy flights.  These correspond to the sticking of a trajectory in phase space to a chaos border.  This results in long durations of near regular behavior, with the orbital period undergoing only small fluctuations.  These are interrupted when the trajectory leaves the border vicinity due to a chaotic dynamical event, and the system re-enters a chaotic state.  Here, the orbital period tends to change significantly with each close encounter.  Ultimately, the tail of the lifetime distribution is thought to be due to Hamiltonian intermittency, with the system switching back and forth between chaotic and close-to-regular, called "quasi-stable", behavior \citep{shevchenko10}.  

There exists an abundance of work throughout the literature on chaos in the three-body problem.  And yet, very few studies have gone beyond the N $=$ 3 limit to consider additional particles \citep[e.g.][]{fregeau04,leigh12,leigh13,leigh15,antognini15}.  
In star clusters, two-body relaxation generally dominates the long-term evaporation of the system.  The collective effects of many long-range gravitational encounters drive diffusion in energy-space, pushing stars into the tail of the Maxwellian velocity distribution and above the escape velocity.  After enough time has passed, the cluster reaches the final stages of dissolution, at which point the number of particles becomes sufficiently small that 
this statistical description of relaxation completely breaks down, and the dominant mode of evolution is entirely 
through close few-body encounters.
If every star in the Universe was born in a star cluster, it follows that every star (and, by extension, every multiple star system) must have passed through such a phase of cluster dissolution (unless, of course, the star in question is still in a cluster at the present epoch).

In recent years, a new paradigm for star cluster evolution has emerged in which low-mass open clusters are, in many ways, more dynamically active than their higher mass globular cluster counter-parts, per unit cluster mass \citep[e.g.][]{leigh11,leigh12,leigh13,leigh13b,geller13,geller15}.  This can be understood by drawing an analogy to the temperature dependence of the composition of a gas, at the atomic and/or molecular level.  More massive clusters (such as globular clusters) have large velocity dispersions and are hence dynamically hot.  This leads to the dissociation or disruption of most multiple star systems during direct dynamical encounters, in analogy with the dissociation or destruction of molecules (and/or atomic ionization) in a hot gas.  Low-mass clusters (such as open clusters), on the other hand, are dynamically cold.  Here, stars and multiples are able to more effectively ``stick together'' during encounters, much like atoms forming molecules in a cold gas.  The longer-lived dynamical interactions produce a more complicated dynamical environment, and contribute to a higher probability for the formation of ``mini-clusters''; tightly bound configurations of a few stars all interacting gravitationally \citep{geller15}.  This increases the probability (per unit cluster mass) for the formation of exotic stellar populations formed from collisions and binary evolution within close multiples.  Perhaps more importantly, the larger-N "mini-clusters" that preferentially and ubiquitously build-up within low-mass open clusters must ultimately dissolve or disrupt to form (both stable and unstable) daughter multiple star systems.  

Indeed, recent observational studies have confirmed the existence of triples in low-mass open clusters in non-negligible numbers \citep{leigh13}, and that these triples could be undergoing dynamical encounters as often as single and binary stars \citep{leigh11}.  This is due to their larger geometrical cross-sections, which renders triples as dynamically active as either single or binary stars in spite of their low number fractions \citep{leigh13}.  Therefore, it seems natural to extend previous statistical studies of the three- and four-body problems in chaotic Newtonian dynamics to larger particle numbers, which has never before been done.

The transition between the regimes where statistical representations of collisional dynamics (e.g., relaxation arguments) are appropriate and where close few-body encounters dominate
is not well understood.  At large N, the relaxation time $\tau_{\rm r}$ (i.e. the time required for the accumulation of small energy changes due to distant encounters to become comparable to the mean energy per star) tends to be much longer than the dynamical or crossing time $\tau_{\rm cr}$ of the system.  But, at small N, these two time-scales are roughly equal \citep[e.g.][]{smith81}.  And yet, the general relation $\tau_{\rm r}/\tau_{\rm cr}$ $\propto$ N/$\log$N, originally derived in the limit of infinite N, remains valid all the way down to N $=$ 16 \citep{mcmillan87}.  Where exactly does the cross-over between these two modes (i.e. relaxation-dominated and strong encounter-dominated)  of disruption occur?  At what critical particle number N?  In this paper, we begin down the long road toward addressing these questions.  Specifically, we go beyond the three-body problem, to consider the decay of interactions involving up to six particles.  We study the distribution of disruption times as a function of the number of interacting particles, for different values of the virial coefficient.  We focus first on encounters involving particles with identical masses, but will go on to consider different mass functions in a forthcoming paper.  

The key idea behind this paper is that we are, in effect, studying the final stages of the cluster dissolution process, but in reverse (i.e. beginning at the lowest particle numbers and working up to higher N).  This approach renders possible a statistical study of the dissolution process using computational or simulation-based methods, since the computer integration times for the simulations scale as N$^2$.  Thus, thousands or even millions of simulations can be performed for a given particle number N and a given set of initial conditions.  This is a critical step toward understanding the origins of the relative numbers of single, binary, triple, etc. stars in the field of our Galaxy, in addition to their orbital parameter distributions, since most (and possibly all) such objects must have at some point passed through a phase of cluster dissolution.  In Section~\ref{method}, we describe the numerical scattering experiments used in this study, which are performed using the \texttt{FEWBODY} code \citep{fregeau04,leigh12}.  Our results are presented in Section~\ref{results}, along with a discussion of their implications in Section~\ref{discussion}.  We summarize our key findings in Section~\ref{summary}.
 
\section{Method} \label{method}

In this section, we present the numerical scattering experiments used to generate and study the distributions of disruption times, as well as the statistical method used to obtain fits to the subsequent distributions.  This is the first paper in a series that aims to go beyond the chaotic three- and four-body problems to consider larger particle numbers.  We use a scattering-based approach in this paper to generate the initial conditions, in agreement with most previous studies in the literature that considered the lifetime statistics of small-number systems.  That is, the initial conditions are highly hierarchical, with large ratios between the orbital separations of the particles.  However, in future studies, we intend to also consider an initially randomized non-hiererarchical state, in which every particle begins with roughly the same energy per unit mass \citep[e.g.][]{mikkola07}.

\subsection{Numerical scattering experiments} \label{experiments}  

We calculate the outcomes of a series of single-binary (1+2; N $=$ 3), binary-binary (2+2; N $=$ 4), single-triple (1+3; N $=$ 4), binary-triple (2+3; N $=$ 5), and triple-triple (3+3; N $=$ 6) encounters using the \texttt{FEWBODY} numerical scattering code\footnote{For the source code, see
http://fewbody.sourceforge.net}.  The code integrates the usual $N$-body equations in configuration- (i.e., position-) space in order to advance the system forward in time, using the eighth-order Runge-Kutta Prince-Dormand integration method with ninth-order error estimate and adaptive time-step.  For more details about the \texttt{FEWBODY} code, we refer the reader to \citet{fregeau04}.  In \citet{leigh12}, we adapted the \texttt{FEWBODY} code to handle encounters involving not only 
single and binary stars, but also triples.\footnote{Specifically we created additional 
subroutines to simulate 1+3 and 3+3 encounters; codes to simulate encounters between 
binaries and singles only, as well as a 2+3 encounter code, were previously available 
in the \texttt{FEWBODY} package.}  

We use the same criteria as \citet{fregeau04} to decide when a given encounter is complete.  To first order, this is defined as the point at which the separately bound hierarchies that make up the system are no longer interacting with each other or evolving internally.  More specifically, the integration is terminated when the top-level hierarchies have positive relative velocity and the corresponding top-level N-body system has positive total energy.  Each hierarchy must also be dynamically stable and experience a tidal perturbation from other nodes within the same hierarchy that is less than the critical value adopted by \texttt{FEWBODY}, called the tidal tolerance parameter.  There is an initial drop-in time that remains constant for a given total encounter energy and angular momentum, which is ultimately set by the tidal tolerance parameter.  For our simulations, this initial time ranges from a few to a few hundred crossing times (see Equation~\ref{eqn:tau_cr}).  Every encounter is terminated precisely one time-step after the criteria described above for stopping the integration are satisfied.  This last time-step is required to ensure that the final hierarchies are indeed dynamically stable.

We adopt the default tidal tolerance parameter in \texttt{FEWBODY} (i.e. $\delta =$ 10$^{-5}$) for all simulations in this paper.  This is chosen to ensure consistency between simulations involving different particle numbers, and reasonable computer integration times.  We note that this could lead to a slight over-estimate of the total encounter lifetimes (see \citet{geller15} for more details).  Importantly, however, there is no reason to expect any systematic dependence of this effect on particle number \citep[e.g.][]{portegieszwart14,geller15}.

We consider three different values for the virial coefficient, namely k $=$ 0, 0.5 and 0.9.  These correspond to initial relative velocities at infinity of 0, 0.5 and 0.9, respectively, in units of the critical velocity v$_{\rm crit}$, defined as the relative velocity at infinity corresponding to a total encounter energy of zero.  For each combination of the virial coefficient k and the number of interacting particles N, we perform 10$^4$ numerical scattering experiments, keeping the orbital semi-major axes and eccentricities constant but varying at random all angles relative to the encounter (the exact number depends on N).  

All orbits are circular, and have semi-major axes of either a$_{\rm 0} =$ 1 AU or a$_{\rm 1} =$ 10 AU.  For example, for a 2+3 encounter, we adopt a$_{\rm 0} =$ 1 AU for the semi-major axes of both the inner orbit of the triple and the interloping binary, and a$_{\rm 1} =$ 10 AU for the outer orbit of the triple.  For a 1+3 encounter, we adopt a$_{\rm 0} =$ 1 AU and a$_{\rm 1} =$ 10 AU for the inner and outer orbits of the triple, respectively, while for a 2+2 encounter we adopt a$_{\rm 0} =$ 1 AU and a$_{\rm 1} =$ 10 AU for the two binaries.  These initial semi-major axes are summarized in Table~\ref{table:initial-conditions}.  A semi-colon separates different objects, and a comma separates the orbits within triples.  Parentheses enclose the semi-major axes of triples, with the smaller of the two separations always corresponding to the inner binary.  Finally, we fix the impact parameter at b $=$ 0 for all simulations.

\begin{table*}
\caption{Initial semi-major axes of all binaries and triples}
\begin{tabular}{|c|c|}
\hline
Encounter Type   &      Semi-major axes      \\
                 &         (in AU)           \\
\hline
1+2              &             1             \\
2+2              &         1; 10           \\
1+3              &        (1, 10)          \\
2+3              &     10; (1, 10)       \\
3+3              &  (1, 10); (1, 10)   \\
\hline
\end{tabular}
\label{table:initial-conditions}
\end{table*}

\subsection{Fitting the distributions of disruption times} \label{fitting}

First, we define the dimensionless time-scale $\tau = \tau_{\rm d}/\tau_{\rm cr}$.  Here, $\tau$ is the disruption time $\tau_{\rm d}$ in units of the crossing time $\tau_{\rm cr}$, or \citep{valtonen06}:  
\begin{equation}
\label{eqn:tau_cr}
\tau_{\rm cr} = \frac{GM^{5/2}}{(2E_{\rm 0})^{3/2}},
\end{equation}  
where M and E$_{\rm 0}$ are the total system mass and energy, respectively.   

To perform the fitting to our disruption time distributions, we adopt a three-parameter function of the form:
\begin{equation}
\label{eqn:func}
f(\tau) = \alpha{\rm e}^{(\tau-\tau_{\rm 0,a})/\tau_{\rm 1/2,a}} + {\rm coth}((\tau-\tau_{\rm 0,b})/\tau_{\rm 1/2,b}) + {\beta},
\end{equation}
where $\alpha$, $\beta$, $\tau_{\rm 0,a}$, $\tau_{\rm 0,b}$, $\tau_{\rm 1/2,a}$ and $\tau_{\rm 1/2,b}$ are all free parameters.  In Equation~\ref{eqn:func}, f($\tau$) is the differential distribution of disruption times (i.e. the cumulative fraction).  That is, the fraction of lifetimes in the interval ($\tau_{\rm d}$,$\tau_{\rm d}$+${\Delta}\tau_{\rm d}$).  For all values of the virial coefficient and particle number, we adopt ${\Delta}\tau_{\rm d} =$ 20$\tau_{\rm cr}$, but note that our results are not sensitive to this choice.  This sets the bin size for the 
histograms (i.e. the 
distribution of disruption times, fit by Equation~\ref{eqn:func}) shown in Figures~\ref{fig:fig1}, ~\ref{fig:fig2} and~\ref{fig:fig3}.  
The disruption times obtained directly from our numerical scattering experiments are normalized by dividing by the total number of simulations performed, ensuring that 0 $\le$ f($\tau$) $\le$ 1.  
We fit Equation~\ref{eqn:func} to the simulated data for every distribution of disruption times, using a Bayesian Markov Chain Monte Carlo (MCMC) approach 
(using the Python \texttt{emcee} package \citep{foreman13}).  
We assume flat priors and Gaussian errors in constructing the likelihood function.  The MCMC method returns a distribution of each parameter, generally,
peaked at the best-fitting value.   We take the median as the ``best-fit'' value, and estimate the 1$\sigma$ uncertainties on each parameter (e.g., see results in Table~\ref{table:one}).

Equation~\ref{eqn:func} is chosen to account for both aspects of the Hamiltonian intermittency characteristic of the chaotic three-body problem which, as we will show, seems to also apply to larger-N interactions.  Equation~\ref{eqn:func} can be thought of as consisting of three parts: the first term (i.e. the exponential term; see Section~\ref{results} for more details) accounts for the exponential decay due to chaotic events stimulating large changes in the orbital period, whereas the third term (i.e. the constant $\beta$) accounts for the hugging of chaos borders characteristic of L\'{e}vy flights.  The second term (i.e. the hyperbolic cotangent function), serves as a bridge between these two limiting functions.  Without the hyperbolic cotangent function, Equation~\ref{eqn:func} transitions too sharply to the long-lived tail and does not provide a good fit to the data.  These contributions to f($\tau$) seem to be quantified by the half-life coefficients $\tau_{\rm 1/2,a}$ and $\tau_{\rm 1/2,b}$ and the constant $\beta$, respectively.  
The constants $\tau_{\rm 0,a}$ and $\tau_{\rm 0,b}$ serve to correct for the arbitrary initial delay in the simulations during which time the two interacting objects drift in from "infinity".  In our simulations, this initial delay time ranges from a few to a few hundred crossing times.  

The main reason behind our choice for the fitting function in Equation~\ref{eqn:func} can be understood as follows.  We begin by trying to fit to the data a functional form for f($\tau$) that consists of an initial exponential drop plus a power-law tail at long encounter durations.  This is motivated by recent work on the lifetime statistics of chaotic three-body interactions  \citep[e.g.][]{shevchenko10,orlov10}.  This can work reasonably well for the three-body case, as has been found in previous studies, but becomes a worse description of the data at larger N.  The reason for this is that the transition from the initial exponential drop to the power-law tail in the chaotic three-body problem is very sharp Ð forming a near 90$^{\circ}$ angle where these two functions intersect.  At larger N, however, this sharp transition is increasingly smoothed out, and some other term is needed in to obtain the required agreement between our choice of fitting function and the data.  We tried many possible functions for f($\tau$), and eventually settled on Equation~\ref{eqn:func}, which is a compromise between the quality of the agreement between our model and the data and minimizing the number of free parameters in our fitting function.

\section{Results} \label{results}

In this section, we present the calculated distributions of disruption times, as well the best-fitting parameter values.

\begin{figure}
\begin{center}
\includegraphics[width=\columnwidth]{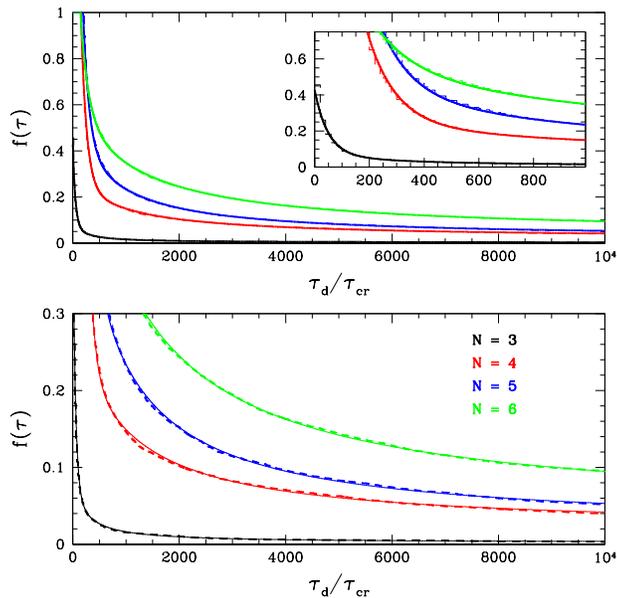}
\end{center}
\caption[Disruption time disruptions for k $=$ 0]{The dashed histograms show the simulated disruption time distributions in units of the crossing time, for a virial coefficient k $=$ 0.  The black, red, blue and green correspond to, respectively, the 1+2 (N $=$ 3), 2+2 (N $=$ 4), 2+3 (N $=$ 5) and 3+3 (N $=$ 6) cases.  In the bottom panel, the (thin) solid lines show the best-fitting parameter values from Equation~\ref{eqn:func} and given in Table~\ref{table:one}.  In the top panel, we also show the disruption time disruptions over the full range in f($\tau$) (i.e. 0 $\le$ f($\tau$) $\le$ 1).  In the inset to the top panel, we show a zoom-in of the transition region where the hyperbolic cotangent term in Equation~\ref{eqn:func} dominates f($\tau$), with the exponential term dominating at smaller $\tau$ and the hyperbolic cotangent term dominating at larger $\tau$.  
\label{fig:fig1}}
\end{figure}

\begin{figure}
\begin{center}
\includegraphics[width=\columnwidth]{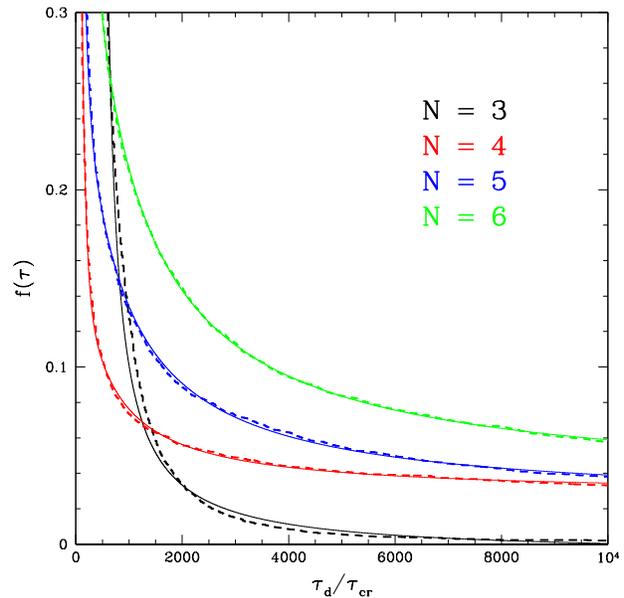}
\end{center}
\caption[Disruption time disruptions for k $=$ 0.5]{The same as in the bottom panel of Figure~\ref{fig:fig1}, but for a virial coefficient k $=$ 0.5.
\label{fig:fig2}}
\end{figure}

\begin{figure}
\begin{center}
\includegraphics[width=\columnwidth]{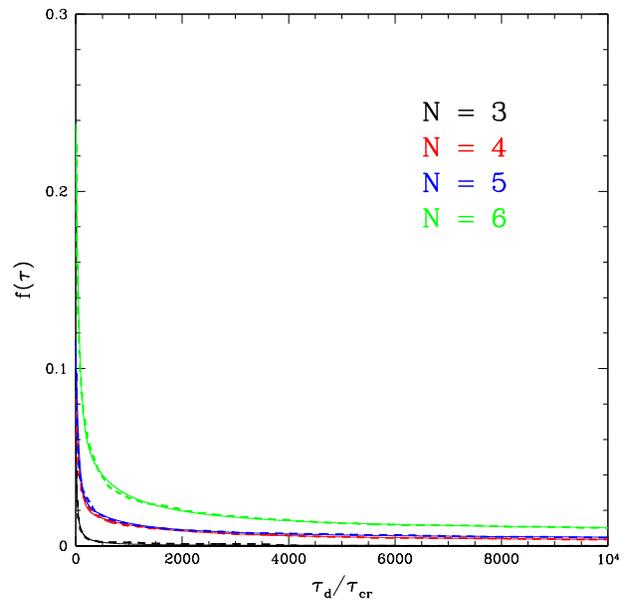}
\end{center}
\caption[Disruption time disruptions for k $=$ 0.9]{The same as in Figure~\ref{fig:fig2}, but for a virial coefficient k $=$ 0.9.
\label{fig:fig3}}
\end{figure}

In Figures~\ref{fig:fig1}, ~\ref{fig:fig2} and~\ref{fig:fig3}, we present the disruption time distributions for virial coefficients of, respectively, k $=$ 0, 0.5 and 0.9.  
As described in the figure captions, the black, red, blue and green lines correspond to the cases, respectively, 1+2 (N $=$ 3), 2+2 (N $=$ 4), 2+3 (N $=$ 5) and 3+3 (N $=$ 6).  The dashed lines show the simulated data, whereas the solid lines show our best fits to these data.  

The best-fit parameter values to Equation~\ref{eqn:func} and their uncertainties are summarized in Table~\ref{table:one}.  Note that for the N $=$ 3 case and the virial ratios k $=$ 0.5 and k $=$ 0.9, we are unable to apply the MCMC method, since all simulations disrupt within a few crossing times
and hence there are too few data points for these simulations to adequately constrain the parameters in Equation~\ref{eqn:func}.
For the k $=$ 0.9 case, which has only a few data points in total, the fit parameters provided in Table~\ref{table:one} are obtained manually (i.e. by eye).  For the k $=$ 0.5 case, however, 
we fit to the data 
only the second two terms in Equation~\ref{eqn:func}, and set $\tau_{\rm 1/2,a}$ $=$ $\tau_{\rm 0,a}$ $=$ 0.  The subsequent best-fitting values for 
 $\tau_{\rm 1/2,b}$, $\tau_{\rm 0,b}$ and $\beta$ are shown in Table~\ref{table:one}.

For both of the first two terms in Equation~\ref{eqn:func}, the system ``half-life'' (i.e. $\tau_{\rm 1/2,a}$ and $\tau_{\rm 1/2,b}$) increases with increasing particle number.  
At least for the first $f(\tau) \gtrsim 0.5$ of the disruption time distributions, the first term in Equation~\ref{eqn:func} corresponding to the initial exponential drop dominates f($\tau$).  Hence, $\tau_{\rm 1/2,a}$ is arguably the more physical half-life than $\tau_{\rm 1/2,b}$ in Equation~\ref{eqn:func}.  More specifically, the $\tau_{\rm 1/2,b}$ derived for Equation~\ref{eqn:func} is too large by several orders of magnitude, relative to the more physical half-life $\tau_{\rm 1/2,a}$.  

Note that the black curve, corresponding to the 1+2 case, intersects with the red curve for the 2+2 case in Figure~\ref{fig:fig2}.  This is due to the arbitrary initial delay in drop-in time, which is long for the 1+2 case for a virial ratio of k $=$ 0.5.  This is compensated for by the constants $\tau_{\rm 0,a}$ and $\tau_{\rm 0,b}$ in Equation~\ref{eqn:func} and shown in Table~\ref{table:one}, as described in Section~\ref{fitting}.

\begin{table*}
\caption{Best-Fit Parameters from Equation~\ref{eqn:func}.  The parameters $\tau_{\rm 1/2, a}$, $\tau_{\rm 1/2,b}$, $\tau_{\rm 0,a}$ and $\tau_{\rm 0,b}$ are given in units of the crossing time (see Equation~\ref{eqn:tau_cr}).}
\begin{tabular}{|c|c|c|c|c|c|c|}
\hline
Particle Number   &  \multicolumn{6}{|c|}{Virial Coefficient k}         \\
\hline
                            &         \multicolumn{6}{|c|}{k $=$ 0}         \\      
\hline
                           &   $\tau_{\rm 1/2,a}$   &   $\tau_{\rm 0,a}$   &   $\tau_{\rm 1/2,b}$   &   $\tau_{\rm 0,b}$   &   $\alpha$     &    $\beta$      \\
                          &                                      &                                   &    ($\times$ 10$^5$)   &                                 &    ($\times$ 10$^{-4}$)   &    ($\times$ 10$^{-4}$)     \\
\hline
1+2; N $=$ 3          &   47.02$^{+75.55}_{-10.54}$   &  90.76$^{+104.36}_{-12.72}$   &   1.666$^{+0.281}_{-0.2863}$   &  -174.73$^{+158.10}_{-176.27}$   &   479.75$^{+259.49}_{-595.02}$   &   19.71$^{+3.92}_{-3.27}$      \\
2+2; N $=$ 4          &   102.57$^{+30.45}_{-10.76}$   &  288.15$^{+35.69}_{-413.37}$    &   23.835$^{+1.698}_{-12.866}$   &  -843.95$^{+853.22}_{-119.56}$   &   1957.52$^{+1433.72}_{-538.73}$   &   197.85$^{+160.47}_{-21.34}$    \\
2+3; N$=$ 5           &  110.14$^{+3.48}_{-5.56}$   &  249.08$^{+31.12}_{-61.98}$    &   34.139$^{+0.660}_{-1.307}$   &  -606.94$^{+50.94}_{-29.03}$   &   3338.61$^{+1671.95}_{-1175.46}$    &   208.89$^{+17.33}_{-8.13}$  \\
3+3; N $=$ 6            &   141.64$^{+7.30}_{-7.10}$   &  219.55$^{+26.07}_{-44.39}$   &   62.756$^{+1.485}_{-1.107}$   &  -1028.91$^{+37.26}_{-39.67}$   &   2492.70$^{+911.18}_{-390.02}$    &   378.25$^{+11.21}_{-15.74}$  \\
\hline
                       &          \multicolumn{6}{|c|}{k $=$ 0.5}          \\
\hline
                           &   $\tau_{\rm 1/2,a}$   &   $\tau_{\rm 0,a}$   &   $\tau_{\rm 1/2,b}$   &   $\tau_{\rm 0,b}$   &   $\alpha$     &    $\beta$      \\
                          &                                      &                                   &    ($\times$ 10$^5$)   &                                 &    ($\times$ 10$^{-4}$)   &    ($\times$ 10$^{-4}$)     \\
\hline            
1+2; N $=$ 3          &   --          &       --         &   6.258$^{+0.009}_{-0.009}$     &  -415.42$^{+0.47}_{-0.43}$   &  --     &     -61.44$^{+0.16}_{-0.19}$  \\
2+2; N $=$ 4       &      64.51$^{+0.99}_{-1.05}$    &  125.48$^{+20.21}_{-18.41}$   &      6.830$^{+0.192}_{-0.178}$   &  -412.02$^{+27.89}_{-28.13}$   &   1342.61$^{+434.17}_{-331.07}$    &  279.21$^{+2.88}_{-2.98}$     \\
2+3; N $=$ 5        &     67.53$^{+1.55}_{-1.73}$    &  140.86$^{+23.97}_{-17.91}$    &   17.583$^{+0.298}_{-0.290}$   &  -572.71$^{+21.23}_{-20.03}$   &   1027.57$^{+278.30}_{-339.68}$     &   223.93$^{+3.92}_{-4.13}$     \\
3+3; N $=$ 6      &     99.33$^{+3.28}_{-3.01}$    &  156.08$^{+44.29}_{-32.95}$    &   29.425$^{+0.331}_{-0.329}$   &  -626.56$^{+16.13}_{-16.33}$   &   1139.75$^{+465.46}_{-405.95}$       &  312.26$^{+4.22}_{-4.22}$     \\ 
\hline                      
                       &           \multicolumn{6}{|c|}{k $=$ 0.9}          \\
\hline
                           &   $\tau_{\rm 1/2,a}$   &   $\tau_{\rm 0,a}$   &   $\tau_{\rm 1/2,b}$   &   $\tau_{\rm 0,b}$   &   $\alpha$     &    $\beta$      \\
                          &                                      &                                   &    ($\times$ 10$^5$)   &                                 &    ($\times$ 10$^{-4}$)   &    ($\times$ 10$^{-4}$)     \\
\hline
1+2; N $=$ 3      &   --          &       --         &     1.00$^{+ --}_{- --}$   &  10.0$^{+ --}_{- --}$   &    --        &     0.0$^{+ --}_{- --}$   \\
2+2; N $=$ 4       &      67.70$^{+6.60}_{-6.66}$    &  -14.61$^{+34.57}_{-42.76}$   &      1.877$^{+0.129}_{-0.103}$   &  -877.74$^{+92.35}_{-106.12}$   &   725.80$^{+644.68}_{-296.88}$    &  18.54$^{+1.28}_{-1.39}$     \\
2+3; N $=$ 5        &     51.57$^{+40.43}_{-10.03}$    &  -46.52$^{+332.76}_{-26.75}$    &   1.372$^{+1.231}_{-0.133}$   &  -500.16$^{+105.09}_{-1090.54}$   &   2094.98$^{+1309.32}_{-2064.35}$     &   35.97$^{+1.87}_{-12.94}$     \\
3+3; N $=$ 6      &     55.74$^{+10.52}_{-7.02}$    &  36.90$^{+116.44}_{-32.39}$    &   2.887$^{+0.225}_{-0.138}$    &   -388.49$^{+49.97}_{-78.96}$   &  806.54$^{+617.68}_{-653.73}$     &  75.38$^{+2.25}_{-3.03}$     \\ 
\hline
\end{tabular}
\label{table:one}
\end{table*}

What characterizes the time evolution of the long-lived encounters?  Based on a visual inspection of our simulations, we see evidence for both long-lived excursions of individual particles as well as the persistence of quasi-stable hierarchies.  For example, during a chaotic N $=$ 5 encounter, one star is often quickly ejected immediately after the encounter begins.  After this, a hierarchical configuration remains consisting of two close binaries in orbit about each other.  This persists as a quasi-stable hierarchy for several orbital periods before decaying back into a chaotic dance.  In fact, to the naked eye, the probability that the system will find itself in a prolonged quasi-stable state immediately after an ejection event appears to be high, although this needs to be verified in future studies.  For example, the initial distributions of energy and angular momenta between the stars, and the corresponding degree of hierarchy in the initial conditions, could be crucial to this issue.


The key features to note in Figures~\ref{fig:fig1}, ~\ref{fig:fig2} and~\ref{fig:fig3} are:  (1) an increase in the system half-lives $\tau_{\rm 1/2,a}$ and $\tau_{\rm 1/2,b}$ with increasing particle number, and (2) an increase in the fraction of long-lived quasi-stable encounters with increasing particle number.  The latter feature can be quantified via an increase in the coefficient $\beta$ (as well as, to a slightly lesser extent, $\tau_{\rm 0,a}$ and $\tau_{\rm 0,b}$) with increasing particle number, for a given virial ratio.  With increasing virial ratio, we also see a clear decrease in the system half-life, as well as an overall decrease in the coefficient $\beta$, corresponding to a lower fraction of long-lived quasi-stable encounters.  

\section{Discussion} \label{discussion}


In this section, we consider the implications of our results for chaotic Newtonian dynamics.  In particular, we wish to formally define the transition between unstable or quasi-stable multiple star systems and star clusters.  
Given our chosen functional form for f($\tau$) in Equation~\ref{eqn:func}, we posit that either of the following transitions at large particle numbers could be used to define the transition of interest:  (1) the system half-lives $\tau_{\rm 1/2,a}$ and $\tau_{\rm 1/2,b}$ asymptote to positive infinity at some critical particle number N$_{\rm crit}$, or (2) the coefficient $\beta$ becomes approximately equal to unity independent of $\tau$ (i.e. yielding f($\tau$) $=$ 1).  Alternatively, the critical particle number N$_{\rm crit}$ marks the point at which Equation~\ref{eqn:func} can no longer describe the disruption time distributions to within some specified tolerance or, more simply, the point at which the initial exponential decay is no longer present.

\begin{figure}
\begin{center}
\includegraphics[width=\columnwidth]{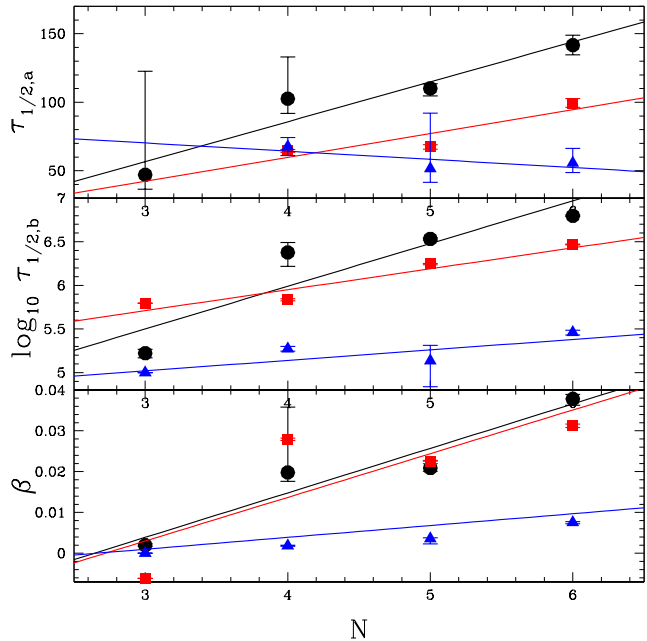}
\end{center}
\caption[Best-fit parameters for Equation~\ref{eqn:func}]{The top, middle and bottom panels show the best-fit values for the parameters $\tau_{\rm 1/2,a}$, (the logarithm of) $\tau_{\rm 1/2,b}$ and $\beta$, respectively, in Equation~\ref{eqn:func}, as provided in Table~\ref{table:one}.  The black circles, red squares and blue triangles correspond to virial coefficients of k $=$ 0, 0.5 and 0.9, respectively.  The solid lines show simple least-squares fits to the data.  
\label{fig:fig4}}
\end{figure}

In Figure~\ref{fig:fig4}, we plot the coefficients $\tau_{\rm 1/2,a}$, $\tau_{\rm 1/2,b}$ and $\beta$ as a function of the number of interacting particles N, for all virial coefficients. 
First, we do not see any evidence for asymptotic behavior in either $\tau_{\rm 1/2,a}$ or $\tau_{\rm 1/2,b}$, at least at such low particle numbers.  
Second, we can attempt an initial estimate of N$_{\rm crit}$, by fitting a line to the data in the bottom panel of Figure~\ref{fig:fig4}, and then 
extrapolating this fit by setting $\beta$ $=$ 1 and solving for N.  
The derived critical particle number N$_{\rm crit}$ depends on the virial coefficient k, at least based on our results for N $<$ 7 (i.e. unless their 
behavior converges at larger N, for example).  
More specifically, we find N$_{\rm crit}$ $\approx$ 89, 91 and 342 for virial coefficients of k $=$ 0, 0.5 and 0.9, respectively, and all equal mass particles.
Thus we estimate that N$_{\rm crit}$ is of order 100 particles. 

More work is needed to explore whether a linear fit is appropriate for $\beta$ as a function of particle number up to N $=$ 100 (and beyond).
We intend to address this issue in further detail in a forthcoming paper, by including additional simulations with more particles ( i.e. N $>$ 6) to 
extend the relations provided in Figure~\ref{fig:fig4} to larger N.
It is also desirable to investigate the dependence of this critical particle number on our assumed encounter parameters, in particular the distribution of 
particle masses.

\section{Summary} \label{summary}

In this paper, we present a possible framework for defining the boundary between stable or quasi-stable multiple star systems and star clusters.  This is done using the distributions of disruption times for chaotic gravitational encounters, which we quantify as a function of the number of interacting particles.  We perform a series of numerical scattering experiments with the \texttt{FEWBODY} code, to calculate the disruption time distributions as a function of both the particle number N and the virial coefficient k.  The subsequent distributions are fit with a physically-motivated function, which consists of two basic parts:  (1) an initial exponential decay followed by (2) a slowly decreasing tail at long encounter times.

The calculated distributions of disruption times show three main features:  (1) the system half-life increases with increasing particle number, (2) the fraction of long-lived quasi-stable encounters increases with increasing particle number and (3) both the system half-life and the fraction of quasi-stable encounters increase with decreasing virial coefficient.  Upon considering the extrapolation of our results to larger-N systems, we suggest that this could potentially offer a clear and unambiguous distinction between star clusters and (stable or quasi-stable) multiple star systems.  Interestingly, our results tentatively suggest that, for all equal mass particles, this transition occurs at a critical particle number of 
order 100 particles, but the exact value depends on the virial coefficient.  


\section*{Acknowledgments}

N.~W.~C.~L. is grateful for the generous support of an NSERC Postdoctoral Fellowship.  
A.~M.~G. is funded by a National Science Foundation Astronomy and Astrophysics Postdoctoral Fellowship under Award No.\ AST-1302765.


\bsp

\label{lastpage}


\begin{thebibliography}{99}

\bibitem[\protect\citeauthoryear{Agekyan \& Anosova}{1967}]{agekyan67} Agekyan T. A., 
Anosova J. P. 1967, Astronomicheskii Zhurnal, 44, 1261
\bibitem[\protect\citeauthoryear{Agekyan \& Anosova}{1983}]{agekyan83} Agekyan T. A., 
Anosova J. P. 1967, Astrophysics, 19, 66
\bibitem[\protect\citeauthoryear{Anosova}{1969}]{anosova69} Anosova J. P. 1969, 
Astrophysics, 5, 81
\bibitem[\protect\citeauthoryear{Anosova \& Orlov}{1983}]{anosova83} Anosova J. P., 
Orlov V. V. 1983, Trudy Astronomical Observatory of Leningrad, 38, 142
\bibitem[\protect\citeauthoryear{Anosova \& Orlov}{1986}]{anosova86} Anosova J. P., 
Orlov V. V. 1986, Soviet Astronomy, 30, 380
\bibitem[\protect\citeauthoryear{Anosova \& Orlov}{1994}]{anosova94} Anosova J. P., 
Orlov V. V. 1994, Celestial Mechanics and Dynamical Astronomy, 59, 327
\bibitem[\protect\citeauthoryear{Antognini \& Thompson}{2015}]{antognini15} Antognini J. M. O., Thompson T. A. 2015, MNRAS, accepted (arXiv:1507.03593)
\bibitem[\protect\citeauthoryear{Foreman-Mackey et al.}{2013}]{foreman13} Foreman-Mackey D., 
Hogg D. W., Lang D., Goodman J. 2013, PASP, 125, 306
\bibitem[\protect\citeauthoryear{Fregeau et al.}{2004}]{fregeau04}
  Fregeau J. M., Cheung P., Portegies Zwart S. F., Rasio F. A. 2004,
  MNRAS, 352, 1
\bibitem[\protect\citeauthoryear{Geller \& Leigh}{2015}]{geller15} Geller A. M., Leigh N. W. C. 2015, ApJL, 808, 25
\bibitem[\protect\citeauthoryear{Geller, Hurley \& Mathieu}{2013}]{geller13} Geller A. M., 
Hurley J. R., Mathieu R. D. 2013, AJ, 145, 8
\bibitem[\protect\citeauthoryear{Leigh, Knigge \& Sills}{2007}]{leigh07} Leigh N.,
Knigge C., Sills A. 2007, ApJ, 661, 210
\bibitem[\protect\citeauthoryear{Leigh \& Sills}{2011}]{leigh11} Leigh N.,
Sills A. 2011, MNRAS, 410, 2370
\bibitem[\protect\citeauthoryear{Leigh \& Geller}{2012}]{leigh12} Leigh N., Geller A. M. 
2012, MNRAS, 425, 2369
\bibitem[\protect\citeauthoryear{Leigh \& Geller}{2013}]{leigh13} Leigh N., Geller A. M.
2013, MNRAS, 432, 2474
\bibitem[\protect\citeauthoryear{Leigh et al.}{2013}]{leigh13b} Leigh N., Giersz M., 
Webb J. J., Hypki A., De Marchi G., Kroupa P., Sills A. 2013, MNRAS, 436, 3399
\bibitem[\protect\citeauthoryear{Leigh \& Geller}{2015}]{leigh15} Leigh N., Geller A. M.
2015, MNRAS, 450, 1724
\bibitem[\protect\citeauthoryear{Leonard}{1989}]{leonard89} Leonard P. J. T. 1989, AJ, 98, 217
\bibitem[\protect\citeauthoryear{Liouville}{1838}]{liouville38}
  Liouville J. 1838, Journ. de Math., 3, 349
\bibitem[\protect\citeauthoryear{McMillan, Hut \& Casertano}{1987}]{mcmillan87} McMillan S. L. W., Hut P., 
Casertano S., 1987, Bulletin of the American Astronomical Society, 19, 715  
\bibitem[\protect\citeauthoryear{Mikkola \& Tanikawa}{2007}]{mikkola07} Mikkola S., Tanikawa K. 2007, MNRAS, 379, L21
\bibitem[\protect\citeauthoryear{Newton}{1686}]{newton1686} Newton,
  I. 1760, Philosophiae Naturalis Principia Mathematica (Trinity
  College:  Regalis Societatis Praesses)
\bibitem[\protect\citeauthoryear{Orlov, Rubinov \& Shevchenko}{2010}]{orlov10} Orlov V. V., Rubinov A. V., Shevchenko I. I. 2010, MNRAS, 408, 1623
\bibitem[\protect\citeauthoryear{Poincare}{1892}]{poincare1892} Poincare H.
1892, Les methodes nouvelles de la mechanique celeste (Paris: Gauthier-Villars)
\bibitem[\protect\citeauthoryear{Portegies Zwart \& Boekholt}{2014}]{portegieszwart14} Portegies Zwart S., Boekholt T. 2014, ApJL, 785, 3
\bibitem[\protect\citeauthoryear{Saslaw, Valtonen \& Aarseth}{1974}]{saslaw74} Saslaw W. C., Valtonen M. J., Aarseth S. J. 1974, ApJ, 190, 253
\bibitem[\protect\citeauthoryear{Shevchenko}{2010}]{shevchenko10} Shevchenko I. I. 2010, Phys. Rev. E, 81, 066216
\bibitem[\protect\citeauthoryear{Smith}{1981}]{smith81} Smith H. Jr. 1981, MNRAS, 195, 35
\bibitem[\protect\citeauthoryear{Szebehely}{1972}]{szebehely72} Szebehely V. 1972, 
Celestial Mechanics, 6, 84
\bibitem[\protect\citeauthoryear{Valtonen}{1975}]{valtonen75} Valtonen M. J. 1975, 
Memoirs of the Royal Astronomical Society, 80, 77
\bibitem[\protect\citeauthoryear{Valtonen \&
    Karttunen}{2006}]{valtonen06} Valtonen M., Karttunen H. 2006, The
  Three-Body Problem (Cambridge: Cambridge University Press)
\end{thebibliography}
\end{document}